# Sovereign risk mitigation mechanism in emerging markets

Ekaterina Bakhmeteva

HSE University, e-mail: esbakhmeteva@hse.ru

Alexey Ponomarenko

Corresponding author, HSE University, e-mail: aaponomarenko@hse.ru

**Abstract**

This paper explores a mechanism for mitigating sovereign risk in emerging markets without risks mutualization. The mechanism involves pooling diversified portfolios of sovereign bonds and issuing them in tranches, with the senior tranche offering low-risk payoffs protected by the subordination of the junior tranches. We argue that this mechanism is feasible for emerging markets. The senior bonds issued by the securitization vehicle attain the properties of a safe asset. The risk level of the junior bonds depends on the structure of the underlying sovereign bonds portfolio. Nevertheless, the properties of the synthetic bonds are, arguably, acceptable for the proposed mechanism's practical application in promoting the development of financial markets in emerging markets and for practical tasks such as intergovernmental aid or lending.

# 1. Introduction

Sovereign risk mitigation mechanisms originally designed for advanced economies hold significant potential for emerging markets (EMs) grappling with financial vulnerabilities. These instruments, which create safe assets without relying on explicit state guarantees, address core challenges like volatile capital flows, banking sector fragility, and limited access to stable funding. Drawing from the analysis of various scholars (e.g. Brunnermeier et al. 2017, Caballero et al. 2016, Leandro and Zettelmeyer 2018, Das et al. 2010) the creation of safe assets aims to achieve several main objectives: first, to create a large and thus liquid bond market; second, to mitigate the “sudden stop” phenomenon experienced by crisis-hit countries; third, to replace sovereign bonds—particularly concentrated exposures to the domestic sovereign—with “safe assets” on the balance sheets of the banking system; and fourth, to increase the global supply of safe assets. These goals, as emphasized by Brunnermeier and Huang (2018), resonate strongly with EM contexts, where sovereign debt crises often amplify economic instability.

The first benefit—creating a large, liquid bond market—is particularly relevant for EMs, where shallow markets exacerbate borrowing costs and volatility. The second aim, mitigating sudden stops, mirrors EM experiences during events like the 1997 Asian Financial Crisis or the 2013 Taper Tantrum, where capital outflows triggered debt distress. By having a stable holder, such as a specialized agency, EMs could buffer against investor flight. Third, replacing concentrated sovereign exposures in banks addresses a vicious "doom loop" prevalent in EMs. The proposed mechanisms could allow EM banks to hold safer senior assets, enhancing financial resilience. Finally, increasing the global supply of safe assets tackles a shortage that Caballero et al. (2016) link to low global interest rates and financial imbalances. EMs could contribute to this supply, attracting safe-haven flows and lowering their own funding costs over time.

Another argument in favor of the relevance of sovereign risk mitigation mechanisms for EMs is intergovernmental aid and lending. For example, China, as a major lender to EMs, has a vested interest in leveraging such risk-free bonds as lending instruments. Over the last few decades, China has extended a substantial volume of loans to EMs, focusing on infrastructure in Africa, Asia, and Latin America. These loans are often collateralized by commodity revenues and routed through Chinese-controlled escrow accounts to minimize default risks (see Gelpern et al. 2025, Horn et al. 2025 for details). If EM borrowers adopted sovereign risk mitigation mechanisms, they could issue low-risk instruments that the lending country could accept as collateral or directly hold. For EM lenders like China, this aligns with strategic motivations: foreign reserves management and exerting global economic influence.

We contribute to this discussion by evaluating the expected properties of the instruments designed to mitigate the sovereign risks of EMs without mutualization. Specifically, we explore a mechanism similar to the ESBies proposed by Brunnermeier et al. (2017) that involves pooling diversified portfolios of sovereign bonds and issuing them in tranches, with the senior tranche offering low-risk payoffs protected by the subordination of junior tranches. We apply this scheme to a cross-section of BRICS+ (BRICS and BRICS partner nations, excluding Belarus and Cuba due to data limitations) EMs.[1] The expected losses of senior and junior bonds under various portfolio compositions are evaluated based on a simulation model.

The rest of the paper is structured as follows. Section 2 outlines the formal model that is used for simulation analysis. Section 3 discusses the design of the sovereign risk mitigation mechanism. Section 4 presents the properties of synthetic senior and junior bonds evaluated based on the simulation results. Section 5 concludes.

## 2. Simulation model

In this section, we outline the general setup and the empirical validation of the model that is used for the simulation analysis.

### 2.1 Business cycle

Sovereign default probabilities increase dramatically during recessions; therefore, we need an algorithm that reproduces the concordance of business cycle states in the EMs.[2] In our model, the probability of a recession ($p_{i,t}$) in country *i* at time *t* is determined as follows:

$$p_{i,t} = \left(1 + \exp\left(F_t + \varepsilon_{i,t}\right)\right)^{-1} \qquad (1)$$

where $F_t$ and $\varepsilon_{i,t}$ are randomly determined and represent the global EM factor and country-specific factors respectively. We set $F_t \sim N(3, 1.9)$ and $\varepsilon_{i,t} \sim N(0, 0.15)$. These parameters allow us to replicate a number of empirical stylized facts.

Specifically, we examine the following indicators: the recession rate (the ratio of recession occurrences to the total number of observations); the average pairwise recession concordance rate (the ratio of coinciding business cycle states across pairs of countries in the cross-section); and the share of variance explained by the first common factor based on the principal component analysis (PCA). Considering that recession occurrence is a discrete variable we report the results for canonical and tetrachoric PCA approaches (see Kolenikov and Angeles 2009 for a discussion).

[1] The choice of countries may be regarded as arbitrary and illustrative. The reasons behind establishing the sovereign default mitigation mechanisms are economic rather than political. Nevertheless, the political connections between the BRICS+ countries, arguably, reinforce the feasibility of such a project. The number of BRICS+ countries also corresponds to the number of countries in the euro area, facilitating the comparison of the calculation results with those of Brunnermeier et al. (2017).

[2] In the model for the euro area, Brunnermeier et al. (2017) assume perfect synchronization of recession occurrences. We believe that this assumption is not realistic for EMs, where business cycles are not closely synchronized (Male 2011). Nonetheless, we have performed the estimations under the perfect synchronization assumption, and the results of our analysis were not dramatically altered.

The empirical estimates are obtained using the annual data on recessions in the BRICS+ countries for the period from 1991 to 2024. All data sources used in the paper are reported in Table A1 in the Appendix.

The model-based estimates were obtained by generating 100,000 samples of the same size as the empirical sample and calculating the statistical measures for each of the simulation runs. The median values of these estimates are reported in Table 1. These results indicate that the degree of recession synchronization implied by the model is consistent with the empirical data.

Table 1. Recession synchronization measures

| | **Recession rate** | **Concordance rate** | **PCA** | **PCA (tetrachoric)** |
|---|---|---|---|---|
| **Empirical** | 0.11 | 0.83 | 0.36 | 0.31 |
| **Model based** | 0.12 | 0.84 | 0.35 | 0.3 |

## 2.2 Sovereign defaults

We proceed by calibrating the sovereign default probabilities (PD) and loss-given-default (LGD) rates for the economies in our cross-section. Following Brunnermeier et al. (2017), we assume that PDs and LGDs are substantially higher in a recession. The resulting parameters are reported in Table 2.

Table 2. Sovereign default model parameters (%)

| | PD1 | PD2 (recession) | LGD1 | LGD2 (recession) |
|---|---|---|---|---|
| **UAE** | 0.75 | 3.5 | 32.5 | 65.0 |
| **China** | 0.75 | 3.5 | 32.5 | 65.0 |
| **Thailand** | 1 | 3.75 | 32.5 | 65.0 |
| **Malaysia** | 1 | 3.75 | 32.5 | 65.0 |
| **India** | 1.5 | 3.75 | 32.5 | 65.0 |
| **Saudi Arabia** | 2 | 4.5 | 32.5 | 65.0 |
| **Indonesia** | 2 | 4.75 | 37.5 | 75.0 |
| **Kazakhstan** | 2 | 5 | 37.5 | 75.0 |
| **Russia** | 2.5 | 5 | 35.0 | 70.0 |
| **Brazil** | 3 | 7 | 50.0 | 80.0 |
| **South Africa** | 3 | 7 | 50.0 | 80.0 |
| **Uzbekistan** | 3 | 7.25 | 60.0 | 80.0 |
| **Egypt** | 3 | 7.25 | 75.0 | 100.0 |
| **Iran** | 3.5 | 7.5 | 75.0 | 100.0 |
| **Nigeria** | 3.5 | 7.75 | 75.0 | 100.0 |
| **Uganda** | 5 | 10 | 75.0 | 100.0 |
| **Ethiopia** | 5.5 | 10 | 100.0 | 100.0 |
| **Bolivia** | 7.5 | 12.5 | 100.0 | 100.0 |

We validate this parametrization by comparing the implied credit default swap (CDS) spreads implied by the model with empirically observed values. For countries without directly available data (Russia and Iran), CDS spreads were estimated based on their CCXI credit ratings, which were mapped to their Moody's and S&P equivalents, using spreads from countries with comparable ratings. The model-based CDS spreads are computed using a simple approximation (see e.g. Hull and White 2001) as follows.

Employing the algorithms described in Section 2.2 and 2.3, we simulate the default occurrences for 10-year sovereign bonds (assuming a 10% coupon) for the countries in our cross-section. Specifically we generate 100000 simulation runs, each run representing the performance of one cohort of sovereign bonds. We calculate country-specific, implicitly observed default rate (*PD*) and loss-given-default rate (*LGD*). The annualized hazard rate (λ) is calculated as:

$$\lambda_i = -\ln(1 - PD_i)/T \qquad (2)$$

where T=10 represents bonds' maturity. We approximate the CDS spread (*s*) as:

$$s_i \approx \lambda_i \times LGD_i + R \qquad (3)$$

where R=0.15% (in line with Brunnermeier et al. 2017) represents the additional counterparty credit risk and liquidity premia.

The results are summarized in Table 3. Based on the obtained CDS spreads, we conclude that our parametrization yields realistic results and the model may be regarded as a valid representation of existing sovereign risks.

In Table 3, we also report the GDP-based portfolio weights for the sovereign bonds of the respective countries and the cumulative expected loss rate for 5-year maturity sovereign bonds. These numbers are used to facilitate the discussion in Sections 3 and 4. The countries in Table 3 are ranked according to their sovereign risks, starting from the safest.

Table 3. Risk properties of sovereign bonds and GDP-based portfolio weights (%)

| | **CDS spread (empirical)** | **CDS spread (model-based)** | **Expected loss rate (model-based)** | **GDP-based weights** |
|---|---|---|---|---|
| **UAE** | 0.6 | 0.6 | 2.5 | 1.5 |
| **China** | 0.6 | 0.6 | 2.4 | 57.2 |
| **Thailand** | 0.7 | 0.8 | 2.6 | 1.7 |
| **Malaysia** | 0.7 | 0.7 | 2.6 | 1.3 |
| **India** | 0.8 | 0.9 | 3.4 | 10.8 |
| **Saudi Arabia** | 1.1 | 1.1 | 5.1 | 3.5 |
| **Indonesia** | 1.2 | 1.3 | 5.5 | 4.1 |
| **Kazakhstan** | 1.3 | 1.3 | 6.0 | 0.7 |
| **Russia** | 1.4 | 1.4 | 6.5 | 6.3 |
| **Brazil** | 2.4 | 2.3 | 11.1 | 6.1 |
| **South Africa** | 2.4 | 2.4 | 11.4 | 1.3 |
| **Uzbekistan** | 3.1 | 3.1 | 14.9 | 0.3 |
| **Egypt** | 3.6 | 3.5 | 17.2 | 1.3 |
| **Iran** | 3.8 | 3.9 | 18.3 | 1.3 |
| **Nigeria** | 3.9 | 4.0 | 18.7 | 1.7 |
| **Uganda** | 5.5 | 5.5 | 26.4 | 0.1 |
| **Ethiopia** | 7.7 | 7.7 | 35.6 | 0.4 |
| **Bolivia** | 10.2 | 10.2 | 45.4 | 0.2 |

## 3. Design of the sovereign risk mitigation mechanism

We analyze the properties of the sovereign risk mitigation mechanism proposed by Brunnermeier et al. (2017). The mechanism works as follows. A specialized entity purchases a diversified portfolio of sovereign bonds, weighted according to a predefined rule. To finance this purchase, the entity issues two types of securities: senior bonds and junior bonds. The tranching point at which the junior tranche is subordinated to the senior tranche is set according to a predefined standard. As a result of this construction, senior and junior bonds are fully collateralized by the underlying portfolio, so that the combined face value of senior and junior bonds equals the sum of the face values of the national sovereign bonds against which they are issued. The resulting balance sheet of the specialized agency is presented in Figure 1.

Figure 1. Balance sheet of the safe EM bonds securitization vehicle

Tranching is key to the safety of senior bonds: losses arising from sovereign defaults would first be borne by holders of the junior bonds; only if they exceed the subordination level, such that the junior bonds are entirely wiped out, would senior bonds begin to suffer any losses. Notably, senior bonds combine the protection due to tranching with the benefit of diversification.

Accordingly, there are two key parameters in this mechanism. The first one is the subordination level (i.e. the share of junior bonds in the specialized agency's liabilities). In our simulation analysis we test various values of this parameter ranging from 0 to 50%.

Another crucial element of the mechanism is the choice of the weight of each country's sovereign bonds in the portfolio. We utilize three approaches to determining portfolio weights.

The first approach employs **GDP-based** weights. This approach is in line with Brunnermeier et al. (2017), who justify this choice with fiscal sustainability considerations. Interestingly, this approach yields a noticeably different portfolio composition of the EM case compared to the euro area case. In the EM cross-section, the resulting share of countries with relatively safe sovereign bonds is higher. Specifically, under the GDP-based weights approach, the sovereign bonds of the top half of the countries (as ranked in Table 3) constitute more than 87% of the portfolio. According to Brunnermeier et al. (2017) in the euro area, the cumulative weight of the half of the countries with safer sovereign bonds is less than 70%.

The second approach (**"two to one"**) implies that sovereign bonds of the relatively safer states constitute two-thirds of the agency's portfolio. Specifically, sovereign bonds of each of the top nine states in Table 3 are assigned a 7.4% weight. The rest of the states are assigned a 3.7% weight.

The third approach applies the "two to one" approach to **China and its debtors**. Under this scheme, the sovereign bonds of China constitute two-thirds of the agency's portfolio. The rest of the portfolio is evenly[3] distributed among the sovereign bonds of its debtor countries from our cross-section: Indonesia, Kazakhstan, Brazil, South Africa, Uzbekistan, Egypt, Uganda, Nigeria, Bolivia, Ethiopia.

The fourth approach employs **equal weights** for sovereign bonds of each country in the cross-section. This approach is used mainly for illustrative purposes.

## 4. Simulation results

We proceed by calculating the risk properties of senior and junior bonds issued under different underlying portfolio structures and subordination levels. Using the model outlined in Section 2, we conduct 100,000 simulation runs, each representing the performance of a cohort of sovereign bonds with 5-year maturity. Based on the results, we calculate the overall expected loss rate of the zero-coupon synthetic senior and junior 5-year bonds issued under the different arrangements. The results are presented in Table 4.

---

[3] For simplicity, we do not reproduce the actual structure of public and publicly guaranteed bilateral credit extended by the Chinese government

The results indicate that senior bonds issued by the specialized securitization agency with an underlying portfolio of EM sovereign bonds can be reliably regarded as a safe asset. Brunnermeier et al. (2017) assume that the bond is safe if its expected loss rate is lower than 0.5%. They show that the subordination level of 0.2 is sufficient to attain this risk level for senior bonds. Our results are generally in line with this finding. Under a subordination level of 0.2–0.25 the expected loss rate of senior bonds is lower than 0.5% for all portfolio structures.[4]

The results for junior bonds are noticeably different. Brunnermeier et al. (2017) show that the expected loss rate for the euro area junior bonds is about 13% at a 0.2 subordination level. This risk level can be attained by EM junior bonds under the GDP-based weights for the underlying portfolio structure (i.e. when the dominant share of the portfolio is constituted by the relatively safe sovereign bonds). Other approaches to forming the underlying portfolio formation do not yield this risk level for junior bonds. Arguably, this result reflects the generally higher level of sovereign risk in EMs compared to that of the euro area countries.

Table 4. Expected loss rates of senior (S) and junior (J) bonds (%)

| | **Portfolio weights choice:** | | | | | | | |
|---|---|---|---|---|---|---|---|---|
| **Subordination** | **GDP-based** | | **Two to one** | | **China and debtors** | | **Equal weights** | |
| **level:** | **S** | **J** | **S** | **J** | **S** | **J** | **S** | **J** |
| **0** | 4.6 | | 10.1 | | 8.1 | | 13.1 | |
| **0.05** | 1.8 | 60.1 | 5.6 | 96.0 | 4.0 | 89.0 | 8.7 | 97.6 |
| **0.1** | 1.0 | 37.9 | 2.0 | 82.3 | 1.8 | 65.9 | 4.5 | 90.0 |
| **0.15** | 0.6 | 27.5 | 0.5 | 64.4 | 1.1 | 48.1 | 1.8 | 77.3 |
| **0.2** | 0.4 | 21.7 | 0.1 | 50.1 | 0.8 | 37.6 | 0.5 | 63.4 |
| **0.25** | 0.2 | 17.9 | 0.0 | 40.3 | 0.5 | 31.0 | 0.1 | 52.0 |
| **0.3** | 0.2 | 15.2 | 0.0 | 33.6 | 0.4 | 26.4 | 0.0 | 43.6 |
| **0.35** | 0.1 | 13.2 | 0.0 | 28.8 | 0.2 | 22.9 | 0.0 | 37.4 |
| **0.4** | 0.0 | 11.7 | 0.0 | 25.2 | 0.2 | 20.2 | 0.0 | 32.7 |
| **0.45** | 0.0 | 10.3 | 0.0 | 22.4 | 0.1 | 17.9 | 0.0 | 29.0 |
| **0.5** | 0.0 | 9.3 | 0.0 | 20.2 | 0.0 | 16.2 | 0.0 | 26.2 |

Nevertheless, we believe that even under alternative portfolio formation approaches, the resulting risk level of junior bonds does not prevent the practical application of the proposed mechanism. We can illustrate the application of the sovereign risk mitigation mechanism with the following example. Suppose that the goal is to replace all public and publicly guaranteed debt owed to the Chinese government of the 10 previously mentioned states with a safe instrument. The total volume of such debt is 32 billion USD. A specialized agency (employing the "China and debtors" approach to portfolio formation under a 0.5 subordination level) is created. It purchases 32 billion USD worth of the debtors' sovereign bonds and 64 billion USD of Chinese sovereign bonds (which is about 19% of the total stock of Chinese foreign debt in bonds). This purchase is funded by the issuance of senior and junior bonds (48 billion USD worth each). The senior bonds (with exceptionally low risk level) may be (partially) purchased by institutions affiliated with China. The issued junior bonds have an expected loss rate of about 16%. This roughly corresponds to the risk level of the Uzbekistan and Egypt sovereign bonds (see Table 3). These bonds are rated BB and B, respectively, by the S&P. The issuance of 48 billion USD worth of such bonds roughly corresponds to about 7.5% of the overall BB - and B - sovereign bonds market. This example is illustrated in Figure 2.

[4] Interestingly, we also found that national-level tranching does not yield risk-free assets even in the case of the safest sovereign bonds.

Figure 2. An illustrative example of a sovereign risk mitigation mechanism implementation

**Before:**

<u>China</u>

| Assets | Liabilities |
|---|---|
| Public and publicly guaranteed credit extended by the Chinese government ($32 bln) | Chinese sovereign bonds ($64 bln) |

<u>Debtor states</u>

| Assets | Liabilities |
|---|---|
| | Public and publicly guaranteed debt to the Chinese government ($32 bln) |

<u>Market investors</u>

| Assets | Liabilities |
|---|---|
| Chinese sovereign bonds ($64 bln) | |

**After:**

<u>China</u>

| Assets | Liabilities |
|---|---|
| Senior bonds ($32 bln) | Chinese sovereign bonds ($64 bln) |

<u>Debtor states</u>

| Assets | Liabilities |
|---|---|
| | Debtor states’ sovereign bonds ($32 bln) |

<u>Market investors</u>

| Assets | Liabilities |
|---|---|
| Senior bonds ($16 bln) | |
| Junior bonds ($48 bln) | |

<u>Securitization vehicle</u>

| Assets | Liabilities |
|---|---|
| Chinese sovereign bonds ($64 bln) | Senior bonds ($48 bln) |
| Debtor states’ sovereign bonds ($32 bln) | Junior bonds ($48 bln) |

## 5. Conclusions

We have explored a sovereign risk mitigation mechanism for a cross-section of EMs. The mechanism involves pooling diversified portfolios of sovereign bonds and issuing them in tranches, with the senior tranche offering low-risk payoffs protected by the subordination of the junior tranches. We argue that this mechanism is feasible for emerging markets. The senior bonds issued by the securitization vehicle attain the properties of a safe asset. The properties of the junior bonds depend on the structure of the underlying sovereign bonds portfolio. The risk level of the junior bonds is relatively low under the GDP-based weights for the underlying portfolio structure; in this case, the dominant share of the portfolio is constituted by the relatively safe sovereign bonds. Under other approaches to portfolio formation, the risk level of the junior bonds is higher. Nevertheless, the risk properties of the junior bonds generally correspond to the B- or BB- rated securities. Arguably, this risk level does not prevent the implementation of the proposed mechanism for promoting the development of financial markets in emerging markets, as well as for practical tasks such as intergovernmental aid or lending.

**DECLARATIONS**

**Availability of data and materials:** The datasets generated during and/or analyzed during the current study are available from the corresponding author upon reasonable request.

**Competing interests:** The authors have no conflicts of interest to declare that are relevant to the content of this article.

**Funding**: The authors did not receive support from any organization for the submitted work.

**Authors' contributions**: Alexey Ponomarenko conceived the presented idea, conducted the calculations, and wrote the manuscript. Ekaterina Bakhmeteva compiled the empirical dataset.

**Acknowledgements:** This article is part of the research project output titled "Approaches to an Alternative International Monetary System", which is associated with the “International Academic Cooperation” project of NRU-HSE. Earlier version of this paper was published as a preprint (Bakhmeteva and Ponomarenko 2026 )

# References

Bakhmeteva, E., Ponomarenko, A. (2026). Sovereign risk mitigation mechanism in emerging markets. arXiv preprint arXiv:2603.22956.

Brunnermeier, M.K., Huang, L. (2018). A global safe asset for and from emerging market economies, Working Paper, 25373, National Bureau of Economic Research.

Brunnermeier, M.K., Langfield, S., Pagano, M., Reis, R., Van Nieuwerburgh, S., Vayanos, D. (2017). ESBies: Safety in the Tranches. Economic Policy, 32 (90), 175-219.

Caballero, R. J., Farhi, E., Gourinchas, P.O. (2016). Safe Asset Scarcity and Aggregate Demand. American Economic Review, 106(5), 513–18.

Das, U.S., Papaioannou, M.G., Trebesch, C. (2010). Sovereign Default Risk and Private Sector Access to Capital in Emerging Markets. IMF Working Paper No. 10/10. Washington: International Monetary Fund.

Gelpern, A., Haddad, O., Horn, S., Kintzinger, P., Parks, B. C., Trebesch, C. (2025). How China Collateralizes. AidData Working Paper No. 136.

Horn, S., Reinhart, C.M., Trebesch, C. (2025). China's Lending to Developing Countries: From Boom to Bust. The Journal of Economic Perspectives, 39(4), 75-100.

Hull, J., White, A. (2001). Valuing Credit Default Swaps II: Modeling Default Correlations. The Journal of Derivatives, 8(3), 12-22.

Kolenikov, S., Angeles, G. (2009). Socioeconomic status measurement with discrete proxy variables: Is principal component analysis a reliable answer? Review of Income and Wealth, 55(1), 128-165.

Leandro, A., Zettelmeyer, J. (2018). The Search for a Euro Area Safe Asset. Working Paper Series WP18-3, Peterson Institute for International Economics.

Male, R. (2011). Developing Country Business Cycles: Characterizing the Cycle. Emerging Markets Finance and Trade, 47(2), 20-39.

# Appendix

Table A1. Statistical Data Sources

| Data | Role in Analysis | Source |
|---|---|---|
| GDP growth (annual %) | Calculating recession synchronization measures | The World Bank. World Development Indicators. GDP growth (annual %). Available at: https://data.worldbank.org/indicator/NY.GDP.MKTP.KD.ZG (accessed: 07.01.2026) |
| Country default spreads | Calibration of PD1, PD2, LGD1, LGD2 parameters | Damodaran A. Country Default Spreads and Risk Premiums. Available at: https://pages.stern.nyu.edu/~adamodar/New_Home_Page/datafile/ctryprem.html (accessed: 09.01.2026) |
| S&P and Moody’s sovereign credit ratings | Approximating of CDS for Russia and Iran | Trading Economics. Credit Rating (S&P, Moody’s, DBRS). Available at: https://tradingeconomics.com/country-list/rating (accessed: 09.01.2026) |
| CCXI sovereign credit rating | Approximating of CDS for Russia and Iran | CCXI. Credit Rating. Available at: https://www.ccxi.com.cn/en/sovereignRating (accessed: 09.01.2026) |
| GDP (current US$) | Calculating GDP-based weights | The World Bank. World Development Indicators. GDP (current US$). Available at: https://data.worldbank.org/indicator/NY.GDP.MKTP.CD (accessed: 07.01.2026) |
| Public and publicly guaranteed debt (from official creditors, bilateral) | Numerical example for the sovereign risk mitigation mechanism | The World Bank. International Debt Statistics (IDS). Available at: https://www.worldbank.org/en/programs/debt-statistics/ids (accessed: 02.02.2026) |
| JPM EMBI Global Diversified sovereign bond index | Estimating the volume of BB and B rated sovereign bonds market | https://www.morningstar.com/etfs/xnas/vwob/portfolio (accessed: 10.02.2026) |